# Gate-dependent Pseudospin Mixing in Graphene/Boron Nitride Moiré Superlattices

Zhiwen Shi[1]*, Chenhao Jin[1]*, Wei Yang[2], Long Ju[1], Jason Horng[1], Xiaobo Lu[2], Hans A. Bechtel[3], Michael C. Martin[3], Deyi Fu[4], Junqiao Wu[4,5], Kenji Watanabe[6], Takashi Taniguchi[6], Yuanbo Zhang[7], Xuedong Bai[2], Enge Wang[8], Guangyu Zhang[2] & Feng Wang[1,5,9]

[1]Department of Physics, University of California at Berkeley, Berkeley, California 94720, USA.

[2]Beijing National Laboratory for Condensed Matter Physics and Institute of Physics, Chinese Academy of Sciences, Beijing 100190, China.

[3]Advanced Light Source Division, Lawrence Berkeley National Laboratory, Berkeley, California 94720, USA.

[4] Department of Materials Science and Engineering, University of California at Berkeley, Berkeley, California 94720, USA.

[5]Materials Science Division, Lawrence Berkeley National Laboratory, Berkeley, California 94720, USA.

[6]Advanced Materials Laboratory, National Institute for Materials Science, 1-1 Namiki, Tsukuba, 305-0044, Japan.

[7]State Key Laboratory of Surface Physics and Department of Physics, Fudan University, Shanghai 200433, China.

[8]International Centre for Quantum Materials, Peking University, Beijing 100871, China.

[9]Kavli Energy NanoSciences Institute at the University of California, Berkeley and the Lawrence Berkeley National Laboratory, Berkeley, California, 94720, USA

*These authors contributed equally to this work.

Electrons in graphene are described by relativistic Dirac-Weyl spinors with a two-component pseudospin[1-12]. The unique pseudospin structure of Dirac electrons leads to emerging phenomena such as the massless Dirac cone[2], anomalous quantum Hall effect[2, 3], and Klein tunneling[4, 5] in graphene. The capability to manipulate electron pseudospin is highly desirable for novel graphene electronics, and it requires precise control to differentiate the two graphene sub-lattices at atomic level. Graphene/boron nitride (graphene/BN) Moiré superlattice, where a fast sub-lattice oscillation due to B-N atoms is superimposed on the slow Moiré period, provides an attractive approach to engineer the electron pseudospin in graphene[13-18]. This unusual Moiré superlattice leads to a spinor potential with unusual hybridization of electron pseudospins, which can be probed directly through infrared spectroscopy because optical transitions are very sensitive to excited state wavefunctions. Here, we perform micro-infrared spectroscopy on graphene/BN heterostructure and demonstrate that the Moiré superlattice potential is dominated by a pseudospin-mixing component analogous to a spatially varying pseudomagnetic field. In addition, we show that the spinor potential depends sensitively on the gate-induced carrier concentration in graphene, indicating a strong renormalization of the spinor potential from electron-electron interactions. Our study offers deeper understanding of graphene pseudospin structure under spinor Moiré potential, as well as exciting opportunities to control pseudospin in two-dimensional heterostructures for novel electronic and photonic nanodevices.

The massless Dirac electrons in graphene are characterized by a unique pseudospin degree of freedom, which exhibit many fascinating transport and optical properties[1-12]. The control of pseudospin, such as opening a pseudospin gap at the Dirac point[19-25], is highly desirable for graphene's application in electronics and photonics. Graphene on atomically flat hexagonal boron nitride (BN) is a promising candidate for pseudospin engineering due to its remarkably high electron mobility[26] and the unique graphene/BN interactions[13-18]. It has been demonstrated recently that new mini-Dirac points and Hofstadter butterfly patterns can emerge from the Moiré superlattice in graphene/BN heterostructures[14-18]. A particularly intriguing property of the Moiré superlattice is that the fast oscillation at B and N sub-lattice sites leads to a periodic spinor potential in graphene that is described by a two-by-two tensor rather than a scalar[27, 28]. This spinor potential couples efficiently to the electron pseudospins, and it was invoked to explain the finite bandgap at the Dirac point in graphene/BN heterostructures[17, 28, 29]. However, direct observation of the spinor potential has been challenging. For example, the density of states change in graphene/BN heterostructures revealed by previous scanning tunneling spectroscopy (STS) and transport measurements can be largely accounted for by a scalar periodic potential[13-18].

Here, we use infrared spectroscopy to probe the spinor potential in the Moiré superlattice. It has been recently predicted that optical conductivity of graphene can exhibit distinctively different behavior in a spinor potential compared to a scalar potential[30]. We demonstrate experimentally that the pseudospin-mixing potential indeed plays a dominant role in optical absorption spectra of graphene/BN heterostructures, owing to the sensitive dependence of optical transition matrix on the hybridized electron wavefunctions. We show that the pseudospin-mixing potential, unlike a scalar potential, can hybridize electron waves with opposite pseudospins and open an "inverse gap" at the boundary of the superlattice Brillouin zone. In addition, we show that the spinor potential depends sensitively on the carrier concentration in graphene, indicating a strong renormalization of the spinor potential from electron-electron interactions.

Our graphene samples were directly grown on hexagonal BN substrate following a van der Waals epitaxy mode[18]. Figure 1a shows an atomic force microscopy (AFM) image of a typical graphene/BN heterostructure, revealing a high coverage of monolayer graphene together with a small portion of bilayer graphene (bright area ~0.3%) and bare BN (dark area ~3%). In the high resolution AFM image (Fig. 1a inset), a triangular Moiré superlattice is clearly observed. The Moiré period of 15±1nm matches well with the lattice constant difference between graphene (2.46 Å) and BN (2.50 Å), suggesting that the epitaxial graphene has a zero lattice twisting angle with BN[18]. Two-terminal field effect graphene devices with back-gate geometry (Fig. 1b and Fig. 2a) were fabricated for electrical and optical characterizations. Figure 1c displays the room temperature transport properties of a typical graphene/BN sample, which exhibits two prominent resistance peaks. The behavior is similar to that observed in previous studies[15-18], where the resistance peak at $V_g = 0$ V and $V_g = -40$ V were attributed respectively to the original Dirac point and the mini-Dirac points on hole side at the m point of the superlattice Brillouin zone (see Fig. 1c inset and Fig. 3a). The resistance peak at the hole side suggests a strong coupling between the zero-twisting graphene and BN layers, and a significant electron-hole asymmetry compared to the much weaker feature at the electron side.

To probe the pseudospin mixing potential from the Moiré superlattice, we performed infrared micro-spectroscopy on graphene/BN heterostructures (Fig. 2a). Figure 2b displays a two-dimensional plot of the transmission spectra difference $T - T_{\text{CNP}}$ at different gate voltages $V_g$ (or equivalently, Fermi energies $E_F$), where $T_{\text{CNP}}$ is transmission spectrum at the charge neutral point (CNP). The Fermi energy is extracted by $E_F = 26.3 \cdot V_g^{1/2}$ (meV) for this sample (Supplementary Information, section 1). The infrared spectra are largely symmetric for electron and hole doping, and show two distinct features: a relatively broad increase of light transmission that systematically shifts to higher energies with increasing $E_F$; and a sharp resonance-like feature at around 380 meV (black dashed line). The broad feature is due to Pauli blocking of interband transitions in bare graphene, which is similar to that observed in graphene on SiO$_2$/Si substrate[7-9]. The sharp feature shows decreased absorption at 380 meV in gated graphene, and is present only in the graphene/BN heterostructure. This energy matches well with the Moiré energy $E_M \equiv \hbar v_F \cdot q_M$ (green arrow in Fig. 1c inset), where $q_M$ is the wavevector of the Moiré pattern and $v_F$ is the graphene Fermi velocity[14]. Therefore, this sharp feature clearly originates from the graphene/BN Moiré superlattice. We note that the Moiré energy happens to be close to the energy of bilayer graphene absorption peak. However, the bilayer graphene will lead to an increased absorption at higher gate voltage[19, 31], which is opposite to the sharp feature observed here. In addition, an estimation of the bilayer signal strength shows that it will be an order of magnitude weaker because of the small percentage of bilayer graphene area in our sample. Figure 2c shows detailed transmission spectra $T - T_{\text{CNP}}$ at several representative electron doping levels that are extracted from horizontal line cuts of Fig. 2b. To better examine the sharp feature associated with Moiré superlattice, we subtract the relative broad background and obtained in Fig. 2d the Moiré superlattice induced optical conductivity change, labeled as $\sigma^M$, around the Moiré energy $E_M$ (Supplementary Information, section 2). Here we have made use of $\sigma^M_{70V} = 0$ because with $V_g = 70$ V ($E_F = 220$ meV) the mini-band optical absorption around $E_M$ is negligible due to Pauli blocking at $2E_F > E_M$. Figure 2d shows a significant absorption peak (corresponding to an increase in optical conductivity) at the Moiré energy $E_M$ for charge neutral graphene. This absorption peak at $E_M$ is opposite from the change in the electron density of state, which shows a prominent dip at $E_M/2$ as observed in previous transport and STS measurements[14-18]. This "inversed" behavior indicates a critical role of optical transition matrix resulting from the unusual electron wavefunction hybridization in graphene on BN. Figure 2d also shows that even slight electron doping can dramatically modify the absorption peak at $E_M$.

To understand the infrared spectra in graphene/BN heterostructures, we need to investigate in detail how the spinor potential from the Moiré superlattice modifies the optical absorption in graphene. A general form of the spinor potential can be written as $V = \sum_{j=1}^{6} V_j e^{i q_j \cdot r}$, where $q_j$ are the reciprocal lattice vectors of the Moiré superlattice with $|q_j| = q_M$[27, 28]. As $q_M$ is much smaller than the valley separation of graphene's original Brillouin zone, the two valleys are effectively decoupled. We can therefore focus on one valley, and determine the other valley by time-reversal symmetry. With the three-fold rotational symmetry and Hermitian requirement, only one among the six $V_j$ is independent, and it can be parameterized with three real numbers $u_0$, $u_3$, $u_1$ as[27, 28]

$$V_1 = V_0 \begin{pmatrix} u_0 + i u_3 & u_1 \\ u_1 & u_0 - i u_3 \end{pmatrix}$$

Here, $V_0$ is a constant characterizing the coupling strength between graphene and the BN substrate. The resulting electron eigen-wavefuction and eigen-energy can be obtained through direct diagnolization of the Hamiltonian $H = \hbar v_F \boldsymbol{\sigma} \cdot \boldsymbol{p} + V$ in the superlattice Brillouin zone. The original Dirac point is at the center γ of the superlattice Brillouin zone, and the high symmetry points at the zone boundary are labeled as m and k/k′, respectively (Fig. 3a).

The three parameters ($u_0$, $u_3$, $u_1$) represent three different types of potential with distinct physical meanings. The $u_0$ term describes a simple scalar potential symmetric at the two sublattices, i.e., a pseudospin-blind potential. The $u_3$ term characterizes the local asymmetry of A-B sublattices, and can be considered as a pseudospin-dependent potential. The off diagonal term $u_1$ mixes the A-B sublattices similar to a pseudo-magnetic field, and can be considered as a pseudospin-mixing potential. These three different types of potential have completely different effects on electron pseudospin, wavefunction hybridization, and optical transitions.

Figure 3b-d shows the optical conductivity changes due to pure $u_0$, $u_3$, and $u_1$ potential with $V_0 = 10 \text{ meV}$, respectively. The insets show the corresponding electronic band dispersion along the γm direction (red line in Fig. 3a) in each case. The optical conductivity change can be best understood by considering electronic states around the m point in the superlattice Brillouin zone (Supplementary Information, section 3 and section 4). The pseudospin-blind potential $u_0$ cannot backscatter Dirac electrons in graphene[4, 5]. Therefore, no gap is opened at the m point and a new mini-Dirac point emerges (Fig. 3b inset). With zero gap at the m point, the effect of a pseudospin-blind potential on the optical absorption is rather small (Fig. 3b). The pseudospin-dependent potential $u_3$ and pseudospin-mixing potential $u_1$, on the other hand, can couple electronic states with opposite pseudospins and both open a nontrivial gap at the m point. However, the hybridized electron wavefunctions at the gapped m point are distinctly different for the $u_3$ and $u_1$ terms, which can be probed directly through optical transitions. For pseudospin-dependent potential $u_3$, only transitions from 1e to 1h and from 2e to 2h sub-band are allowed close to the mini-gap (Fig. 3c inset and Supplementary Information, section 4). In this case, the absorption spectrum mimics the electron density of states[13, 14] except that the energy scale is multiplied by 2, and it shows an absorption dip at 380 meV (Fig. 3c). The mini-gap generated by $u_3$ can therefore be termed as a "normal" gap. The pseudospin-mixing potential $u_1$, on the other hand, restricts the optical transitions to the largely parallel 1e-2h and 2e-1h sub-bands (Fig. 3d inset and Supplementary Information, section 4). Transitions between the parallel bands lead to a van Hove singularity in the joint density of states and to a large absorption peak at $E_M$ (Fig. 3d), opposite to the case in Fig. 3c. We term the mini-gap generated by $u_1$ as an "inverse" gap. Our simulated optical conductivity is also consistent with results in ref. [30].

When $u_0$, $u_1$, and $u_3$ are all finite, their interplay further modifies the electron hybridization and optical spectra. The size of the mini-gap at the m point is described by $u_3 \pm u_1$ for the valence/conduction band. The relative magnitude of $|u_1|$ and $|u_3|$ strongly affects the nature of the mini-gap, which crosses zero for either the valence or conduction band at $|u_1| = |u_3|$. When $|u_1| < |u_3|$, the mini-gap is more similar to a "normal" gap induced by a pure $u_3$ potential, and it leads to an optical absorption dip at $E_M$. On the other hand, the mini-gap is more similar to an "inverse gap" induced by a pure $u_1$ potential when $|u_1| > |u_3|$, which produces an optical absorption peak at $E_M$ (Supplementary Information, section 4). The $u_0$ term does not affect the mini-gap at the m point, and only slightly modify the optical absorption spectra. The observed absorption peak at $E_M$ for charge neutral graphene (Fig. 3e) obviously cannot be described by the $u_0$ scalar potential, and it has a lineshape similar to that produced by the pseudospin-mixing $u_1$ term. It demonstrates unambiguously the spinor potential nature of the Moiré superlattice potential, and shows that the pseudospin-mixing term $u_1$ is the dominating component. Quantitative comparison with the theory shows that the observed absorption spectrum can be described nicely using parameters obtained from a microscopic model (Supplementary Information, section,5) with $V_0 = 10$ meV and $(u_0, u_3, u_1) = (1/2, -\sqrt{3}/2, -1)$ (Fig. 3e). The positive value of $V_0$ arises from a stronger carbon-boron coupling than the carbon-nitrogen coupling, presumably due to the significantly larger radius of the p orbital in boron than in nitrogen (Supplementary Information, section 5). The resulting electronic bandstructure from this set of parameters is displayed in Fig. 3f. It shows a much stronger bandstructure change at the hole side, consistent with the electron-hole asymmetry observed in electrical transport. This asymmetry is not pronounced in the optical data because optical transitions always involve both the electron and hole states.

Next, we examine the gate-dependence of optical absorption spectra around $E_M$. We plotted the peak height at $E_M$ for different Fermi energies in Fig. 4, which shows a sharp decrease with increased electron concentration and almost goes to zero at $E_F \sim 140$ meV. This sensitive dependence on electron doping is quite interesting because it cannot be explained by the single-particle Pauli blocking effect: the relevant Fermi energy is too low to block electronic state transition at the m point (at $E_M/2 = 190$ meV, indicated by the dashed line). Therefore, the decreased absorption peak at $E_M$ should originate from a change in the optical transition matrices, indicating that the spinor potential of the Moiré superlattice is modified appreciably in doped graphene due to electron-electron interactions. It is well known that dielectric screening from free carriers can reduce the scalar electrostatic potential, which can be calculated using the random phase approximation (RPA)[32]. If we assume that the effective spinor potential is screened like the scalar potential with wavevector $q_M$, the RPA calculation predicts a rather weak decrease of the potential and the absorption peak with the carrier doping[32] (orange line in Fig. 4). Obviously the RPA approximation is not applicable to the spinor potential in graphene. Recent studies based on renormalization group theory show that the pseudospin-dependent potential is strongly renormalized by electron-electron interactions[29]. Presumably the spinor potential becomes weaker with electron doping due to such renormalization effects, and the $u_1$ and $u_3$ parameters can have different renormalization behavior. Further theoretical studies need to be carried out to quantitatively describe the experimental data.

In conclusion, we demonstrate infrared spectroscopy as a sensitive probe for the Moiré spinor potential in graphene/BN heterostructure, and determine the dominant effect of an unusual pseudospin-mixing potential. In addition, our data indicates that the spinor potential can be strongly renormalized by carrier doping from electron-electron interactions. More generally, our study shows that nontrivial manipulation of electron pseudospins can be achieved in two dimensional graphene heterostructures. It opens up new avenues in engineering pseudospin in such heterostructures for novel electronic and photonic nanodevices.

**Methods**

Graphene samples were directly grown on hexagonal BN substrate without catalyst following a van der Waals epitaxial mode. The growth was carried out in a remote plasma enhanced chemical vapor deposition (R-PECVD) system at ~500 °C with pure $CH_4$ as carbon source. Hydrogen plasmon etching was used after the growth to etch away the second layer and obtain larger proportion of monolayer graphene. We used TEM grids as shadow masks for metal electrode deposition. A long working distance optical microscope was employed to find BN flakes and align the shadow mask with the chosen BN flake. The deposited metal film is 2nm/80nm Ti/Au. Transmitted infrared spectra were measured using a Fourier transform infrared microscope (Thermo Nicolet Nexus 870 with Continuum XL IR Microscope) with a synchrotron infrared light source. All the measurements were performed in vacuum at room temperature.

**References**

1   Novoselov, K. S. *et al.* Electric field effect in atomically thin carbon films. *Science* 306, 666-669 (2004).
2   Novoselov, K. S. *et al.* Two-dimensional gas of massless Dirac fermions in graphene. *Nature* 438, 197-200 (2005).
3   Zhang, Y. B., Tan, Y. W., Stormer, H. L. & Kim, P. Experimental observation of the quantum Hall effect and Berry's phase in graphene. *Nature* 438, 201-204 (2005).
4   Neto, A. H. C. *et al.* The electronic properties of graphene. *Reviews of Modern Physics* 81, 109 (2009).
5   Katsnelson, M. I., Novoselov, K. S. & Geim, A. K. Chiral tunnelling and the Klein paradox in graphene. *Nature Physics* 2, 620-625 (2006).
6   Geim, A. K. & Novoselov, K. S. The rise of graphene. *Nature Materials* 6, 183-191 (2007).
7   Wang, F. *et al.* Gate-variable optical transitions in graphene. *Science* 320, 206-209 (2008).

8       Li, Z. Q. *et al.* Dirac charge dynamics in graphene by infrared spectroscopy. *Nature Physics* 4, 532-535 (2008).

9       Horng, J. *et al.* Drude conductivity of Dirac fermions in graphene. *Physical Review B* 83, 165113 (2011).

10      Min, H., Borghi, G., Polini, M. & MacDonald, A. H. Pseudospin magnetism in graphene. *Physical Review B* 77, 041407 (2008).

11      Jung, J., Zhang, F. & MacDonald, A. H. Lattice theory of pseudospin ferromagnetism in bilayer graphene: Competing interaction-induced quantum Hall states. *Physical Review B* 83, 115408 (2011).

12      San-Jose, P., Prada, E., McCann, E. & Schomerus, H. Pseudospin Valve in Bilayer Graphene: Towards Graphene-Based Pseudospintronics. *Physical Review Letters* 102, 247204 (2009).

13      Park, C.-H. *et al.* New generation of massless Dirac fermions in graphene under external periodic potentials. *Physical Review Letters* 101, 126804 (2008).

14      Yankowitz, M. *et al.* Emergence of superlattice Dirac points in graphene on hexagonal boron nitride. *Nature Physics* 8, 382-386 (2012).

15      Ponomarenko, L. A. *et al.* Cloning of Dirac fermions in graphene superlattices. *Nature* 497, 594-597 (2013).

16      Dean, C. R. *et al.* Hofstadter's butterfly and the fractal quantum Hall effect in moire superlattices. *Nature* 497, 598-602 (2013).

17      Hunt, B. *et al.* Massive Dirac fermions and Hofstadter butterfly in a van der Waals heterostructure. *Science* 340, 1427-1430 (2013).

18      Yang, W. *et al.* Epitaxial growth of single-domain graphene on hexagonal boron nitride. *Nature Materials* 12, 792-797 (2013).

19      Zhang, Y. B. *et al.* Direct observation of a widely tunable bandgap in bilayer graphene. *Nature* 459, 820-823 (2009).

20      McCann, E. Asymmetry gap in the electronic band structure of bilayer graphene. *Physical Review B* 74, 161403 (2006).

21      Ohta, T. *et al.* Controlling the electronic structure of bilayer graphene. *Science* 313, 951-954 (2006).

22      Lui, C. H. *et al.* Observation of an electrically tunable band gap in trilayer graphene. *Nature Physics* 7, 944-947 (2011).

23      Son, Y. W., Cohen, M. M. & Louie, S. G. Half-metallic graphene nanoribbons. *Nature* 444, 347-349 (2006).

24      Han, M. Y., Ozyilmaz, B., Zhang, Y. & Kim, P. Energy band gap engineering of graphene nanoribbons. *Phys. Rev. Lett.* 98, 206805 (2007).

25      Avouris, P., Chen, Z. H. & Perebeinos, V. Carbon-based electronics. *Nature Nanotechnology* 2, 605-615 (2007).

26      Dean, C. R. *et al.* Boron nitride substrates for high-quality graphene electronics. *Nature Nanotechnology* 5, 722-726 (2010).

27      Wallbank, J. R. *et al.* Generic miniband structure of graphene on a hexagonal substrate. *Physical Review B* 87, 245408 (2013).

28      Kindermann, M., Uchoa, B. & Miller, D. L. Zero-energy modes and gate-tunable gap in graphene on hexagonal boron nitride. *Physical Review B* 86, 115415 (2012).

29      Song, J. C. W., Shytov, A. V. & Levitov, L. S. Electron Interactions and Gap Opening in Graphene Superlattices. *Physical Review Letters* 111, 266801 (2013).
30   Abergel, D. S. L. *et al.* Infrared absorption by graphene-hBN heterostructures. *New Journal of Physics* 15, 123009 (2013).
31   Li, Z. Q. *et al.* Band Structure Asymmetry of Bilayer Graphene Revealed by Infrared Spectroscopy. *Physical Review Letters* 102 (2009).
32   Hwang, E. H. & Das Sarma, S. Dielectric function, screening, and plasmons in two-dimensional graphene. *Physical Review B* 75, 205418 (2007).

**Author Information** The authors declare no competing financial interests. Correspondence and requests for material should be addressed to F.W. (fengwang76@berkeley.edu, for measurement and theory details) and G.Z. (gyzhang@aphy.iphy.ac.cn, for details of sample preparation).

**Figure Captions:**

**Figure 1| Graphene/BN heterostructure and typical transport property. a**, Atomic force microscopy image showing high coverage of monolayer graphene together with a small portion of bilayer graphene (bright area ~0.3%) and bare BN (dark area ~3%). The inset displays a high-resolution AFM image of the graphene/BN Moiré superlattice with a period of $15\pm1$nm. **b**, Optical micrograph of a two-terminal field effect graphene/BN device on a $SiO_2$/Si substrate. **c**, Gate-dependent resistance of a typical graphene/BN device at room temperature. The resistance peak at $V_g = 0$ V and $V_g = -40$ V corresponds to the original Dirac point and the mini-Dirac points on hole side at the m point of the superlattice Brillouin zone, respectively. The inset shows the linear band of graphene. The Moiré wavevector $q_M$ (red arrow) connects the superlattice m point. Optical transitions at the m point have energy $E_M = q_M \cdot v_F$ (green arrow).

**Figure 2 | Infrared micro-spectroscopy of the graphene/BN heterostructure. a**, Schematic drawing of the experimental setup. **b**, Two-dimensional plot of the transmission spectra difference $T - T_{\text{CNP}}$ at different Fermi energies $E_F$, where $T_{\text{CNP}}$ is transmission spectrum for graphene at charge neutral point(CNP). The sharp feature at around 380 meV (black dashed line) originates from the Moiré superlattice. The broad feature that shifts with $E_F$ is due to Pauli blocking of interband transitions. **c**, Transmission spectra $T - T_{\text{CNP}}$ at several representative electron doping levels (Fermi energies and corresponding gate voltages are show in the legend), extracted from horizontal line cuts of **b**. **d**, Moiré superlattice induced optical conductivity change $\sigma^M$ at different gate voltages. $\sigma_0 = \pi e^2/2h$ is graphene universal conductivity.

**Figure 3 | Calculated optical conductivity changes under different spinor potentials. a**, Mini-Brillouin zone of the Moiré superlattice. γ point corresponds to graphene's original Dirac point. Mini-Bruillouin zone boundary and edges are labeled with m and k/k′. Red line indicates the γm direction. **b-d**, Optical conductivity changes at charge neutral point under $u_0$, $u_3$, and $u_1$ potential, respectively. $V_0 = 10$ meV. The insets show corresponding band structures along the γm direction and allowed optical transitions near the m point. **b,** The $u_0$ potential does not open a gap at m point, and has small effect on optical conductivity. **c,** The $u_3$ potential opens a "normal" gap, where optical transitions are restricted to symmetric 1e-1h and 2e-2h bands. This leads to a dip at $E_M$ in optical conductivity. **d,** The $u_1$ potential opens an "inverse" gap where only 1e-2h and 2e-1h transitions are allowed around the m point. Such

transitions between parallel bands lead to an absorption peak at $E_M$ due to a van Hove singularity in joint density of states. **e**, Comparison of experimental and theoretical optical conductivity change using the spinor potential from a microscopic model. **f**, 3D mini-band structure in superlattice Brillouin Zone with the parameters in **e**. The hole side is modulated much stronger than the electron side.

**Figure 4 | Gate-dependent Moiré spinor potential.** The optical conductivity peak at $E_M$ depends sensitively on the electron doping in graphene (blue points), and it diminishes before the optical transitions are affected by Pauli blocking (at the dashed vertical line). It suggests a strong renormalization of the Moiré spinor potential by electron-electron interactions, which cannot be described by simple dielectric screening using the random phase approximation (orange line).

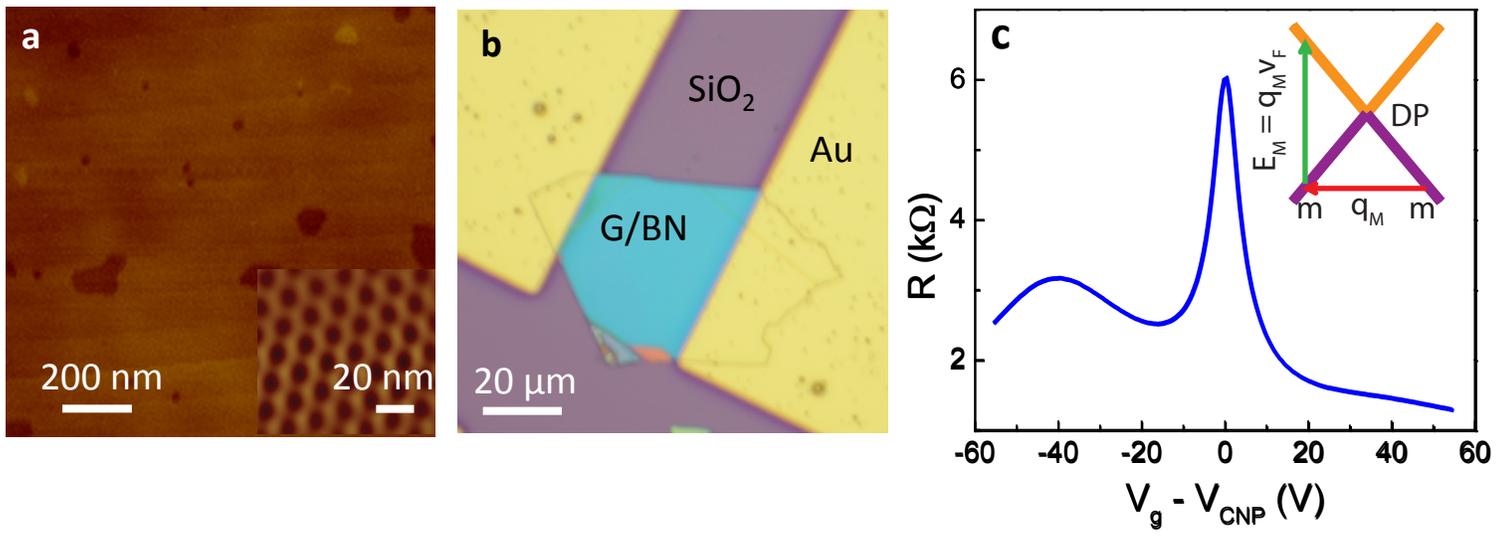

Figure 1

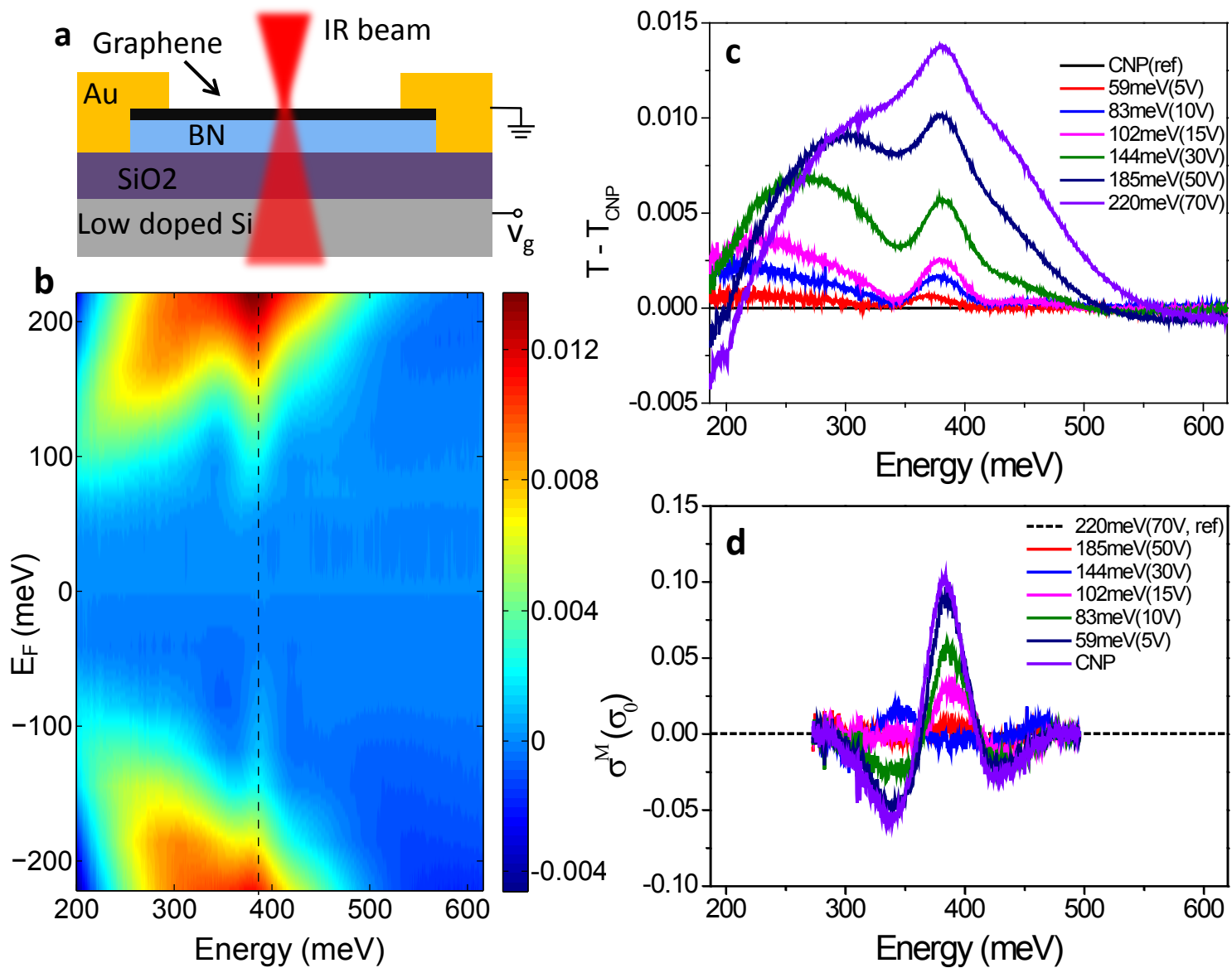

Figure 2

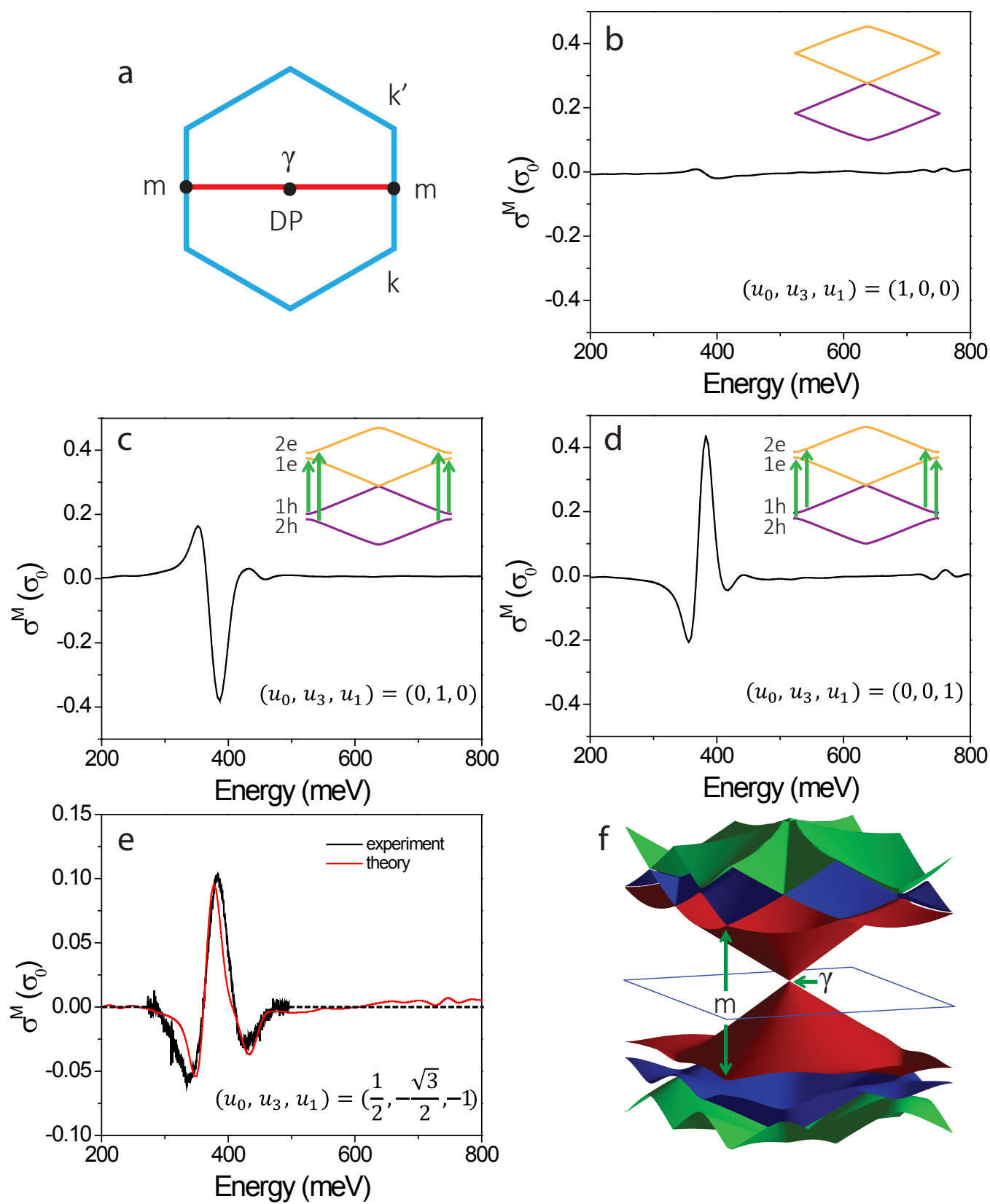

Figure 3

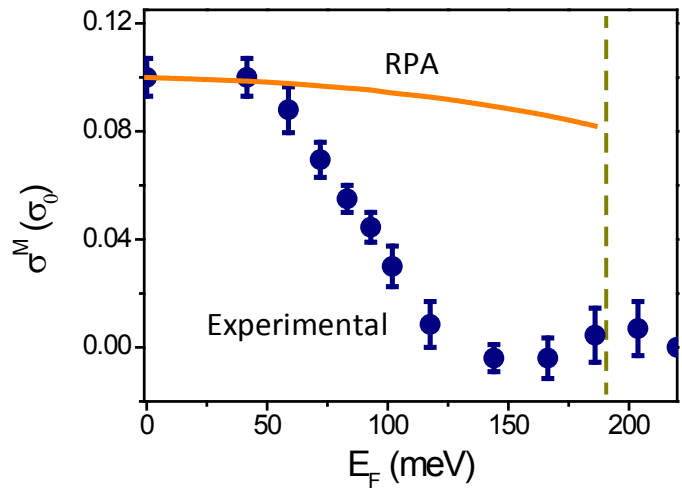

Figure 4

Supplementary information for

# Observation of Pseudospin-mixing Potential in Graphene/Boron Nitride Moiré Superlattice

Zhiwen Shi[1]*, Chenhao Jin[1]*, Wei Yang[2], Long Ju[1], Jason Horng[1], Xiaobo Lu[2], Hans Bechtel[3], Michael C. Martin[3], Deyi Fu[4], Junqiao Wu[4,5], Kenji Watanabe[6], Takashi Taniguchi[6], Yuanbo Zhang[7], Xuedong Bai[2], Enge Wang[8], Guangyu Zhang[2] & Feng Wang[1,5,9]

[1]Department of Physics, University of California at Berkeley, Berkeley, California 94720, USA.

[2]Beijing National Laboratory for Condensed Matter Physics and Institute of Physics, Chinese Academy of Sciences, Beijing 100190, China.

[3]Advanced Light Source Division, Lawrence Berkeley National Laboratory, Berkeley, California 94720, USA.

[4] Department of Materials Science and Engineering, University of California at Berkeley, Berkeley, California 94720, USA.

[5]Materials Science Division, Lawrence Berkeley National Laboratory, Berkeley, California 94720, USA.

[6]Advanced Materials Laboratory, National Institute for Materials Science, 1-1 Namiki, Tsukuba, 305-0044, Japan.

[7]State Key Laboratory of Surface Physics and Department of Physics, Fudan University, Shanghai 200433, China.

[8]International Centre for Quantum Materials, Peking University, Beijing 100871, China.

[9]Kavli Energy NanoSciences Institute at the University of California, Berkeley and the Lawrence Berkeley National Laboratory, Berkeley, California, 94720, USA

*These authors contributed equally to this work.

**S1. Determination of the Fermi energy**

**S2. Extraction of Moiré potential-induced feature**

**S3. Selection rules for optical transitions away from the superlattice Brillouin Zone (sBZ) boundary**

**S4. Selection rules for optical transitions near sBZ boundary**

**S5. Microscopic model to determine $u_0$, $u_3$ and $u_1$**

## S1. Determination of the Fermi energy

We determine here the gate efficiency of the device from which the optical data was obtained. The gate-dependent Fermi level of graphene is calculated using the parallel plate capacitor model. There are two different dielectric materials between the gate electrode and graphene: SiO2 ($d_1$ = 285 nm, $\varepsilon_1$ = 3.90) and BN ($d_2$ = 161 nm, $\varepsilon_2$ = 5.09)[1]. Based on these data, carrier density induced by the backgate voltage is described by $n = 5.1 \times 10^{10} \text{cm}^{-2} * V_g$. In graphene, the electron density is related to Fermi energy as $n = E_F^2/\pi\hbar^2 v_F^2$. We therefore obtain the relation between Fermi energy and gate voltage $E_F = 26.3\sqrt{V_g}$, where $E_F$ in unit of meV and $V_g$ in unit of V.

## S2. Extraction of Moiré potential-induced absorption feature

To extract the absorption feature around $E_M$ induced by the Moiré potential, we decompose the absorption spectra in Fig. 2c into a smooth and broad absorption $T^B$ (Fig. S1a) due to Pauli-blocking of interband transitions in bare graphene with linear band[2,3]; and a sharp feature $T^M$ (Fig. S1b) from the Moiré sueprlattice. From the transmission spectra, we can obtain the corresponding optical conductivity change due to the Moiré potentail $\sigma^M = \left[(T^M - T^M_{70V})/c_l\pi\alpha\right] \cdot \sigma_0$, where $\sigma_0$ is graphene universal conductivity, $c_l$ is the local field factor, and $\alpha$ is the fine structure constant, $c_l\pi\alpha$=2%[4]. Here we use the highly doped graphene spectrum $T^M_{70V}$ (with $E_F$ = 220 meV) as the reference because the mini-band optical conductivity $\sigma^M_{70V} = 0$ due to Pauli blocking with $2E_F > E_M$.

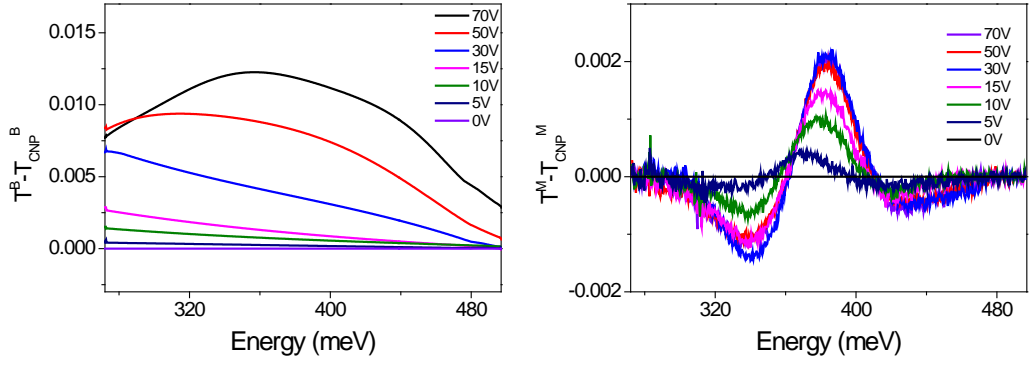

**Figure S1: (a)** Fitted broad feature $T^B - T^B_{CNP}$ due to Pauli-blocking of interband transitions in bare graphene with linear band. **(b)** Extracted sharp feature $T^M - T^M_{CNP}$ around $E_M$ from the Moiré superlattice.

## S3. Selection rules for optical transitions away from superlattice Brillouin Zone (sBZ) boundary

Under the Moiré superlattice spinor potential, graphene electronic states with different wavevectors are perturbed differently: states away from sBZ boundary (region I in Fig. S2a) only couple to other states with very different energies and the coupling can be described using the non-degenerate perturbation theory; states near sBZ boundary (region II in Fig. S2a), on the other hand, have to be described using the degenerate perturbation theory. They give rise to very different selection rules.

For states in region I, the band dispersion is little affected; and optical transitions are allowed only between symmetric valence and conduction bands (green arrow in Fig. S2a), which is similar to that in bare graphene. The other possible transitions (blue arrow in Fig. S2a) are forbidden due to a zero transition dipole moment within the lowest order perturbation, as we

explain below:

Consider one typical transition of this type from the valence state $\psi_a^{v\prime}$ to the conduction state $\psi_b^{c\prime}$. Here a, b labels a state from the second and first subband, respectively. Figure S2b also shows the two states in graphene's original Brillouin zone. It is clear that the two wavevectors $|\mathbf{k}_a - \mathbf{k}_b| = q_M$.

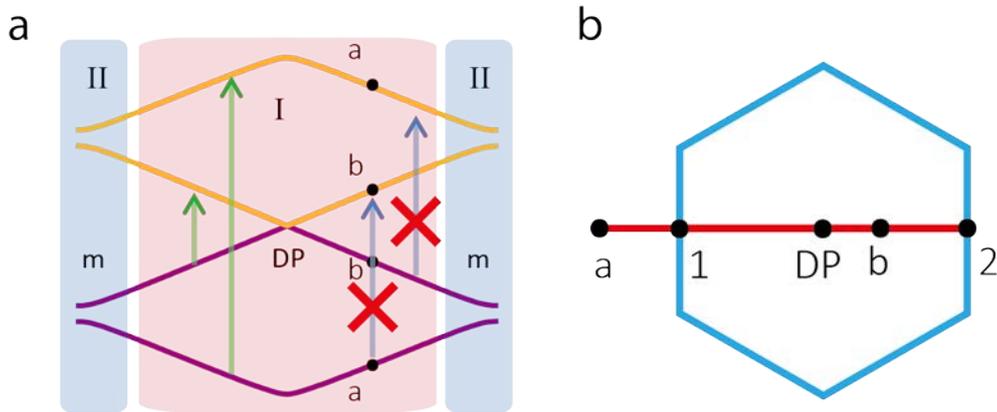

**Figure S2: Schematic drawing of optical selection rules.** **(a)** Electronic coupling between states away from the sBZ boundary (region I, the red shaded area) are described by non-degenerate perturbation theory. Optical transitions in this region are only allowed between symmetric bands, and the resulting absorption is almost identical to that from corresponding transitions in bare graphene. In region II (blue shaded area) the Moiré superlattice potential mixes almost degenerate states, and leads to strong modification in optical absorption. **(b)** Representative coupled states from region I (a and b) and from region II (1 and 2) in the original graphene Brillouin zone. The blue hexagon is the sBZ.

Using the second order perturbation theory, we can obtain the perturbed electronic state wavefunction as

$$\psi_a^{v\prime} = \psi_a^v + \frac{u_0}{-E_M}\psi_b^c = \psi_a^v - C_1\psi_b^c$$

$$\psi_a^{c\prime} = \psi_a^c + \frac{u_0}{E_M}\psi_b^v = \psi_a^v + C_1\psi_b^v$$

$$\psi_b^{v\prime} = \psi_b^v + \frac{u_0}{-E_M}\psi_a^v = \psi_b^v - C_1\psi_a^c$$

$$\psi_b^{c\prime} = \psi_b^c + \frac{u_0}{-E_M}\psi_a^v = \psi_b^c + C_1\psi_a^v \quad (S2)$$

Based on optical transition matrix element in bare graphene[4], we have:

$$\langle\psi_a^v|H_{opt}|\psi_b^v\rangle = \langle\psi_a^v|H_{opt}|\psi_b^c\rangle = \langle\psi_a^c|H_{opt}|\psi_b^v\rangle = -\langle\psi_a^c|H_{opt}|\psi_b^c\rangle = 0 \quad (S3)$$

$$\langle\psi_a^v|H_{opt}|\psi_a^v\rangle = \langle\psi_b^c|H_{opt}|\psi_b^c\rangle = -\langle\psi_a^c|H_{opt}|\psi_a^c\rangle = -\langle\psi_b^v|H_{opt}|\psi_b^v\rangle = -v_F\boldsymbol{E}\cdot\hat{\mathbf{k}}_a \equiv T_1$$

$$\langle\psi_a^v|H_{opt}|\psi_a^c\rangle = \langle\psi_b^c|H_{opt}|\psi_b^v\rangle = -\langle\psi_a^c|H_{opt}|\psi_a^v\rangle = -\langle\psi_b^v|H_{opt}|\psi_b^c\rangle = iv_F\hat{\mathbf{z}}\cdot(\boldsymbol{E}\times\hat{\mathbf{k}}_a) \equiv T_2$$

Here $H_{opt} = \boldsymbol{E}\cdot\boldsymbol{\nabla}$ is the optical transition Hamiltonian, $\boldsymbol{E}$ is the light electric field, $\hat{\mathbf{k}}_a$ is the unit vector along $\gamma a$, $\hat{\mathbf{z}}$ is the unit vector along z direction perpendicular to graphene plane. As equation S3 shows, in unperturbed bare graphene optical transitions between $\psi_a^v$ and $\psi_b^c$ are not allowed because $\mathbf{k}_a \neq \mathbf{k}_b$. This constraint is possibly relaxed by the Moiré potential because it mixes electronic states with different momentum. However, our calculation shows that optical transition between $\psi_a^{v\prime}$ and $\psi_b^{c\prime}$ is still forbidden because the transition matrix element is:

$$\langle\psi_a^{v\prime}|H_{opt}|\psi_b^{c\prime}\rangle = C_1\left(\langle\psi_a^v|H_{opt}|\psi_a^v\rangle - \langle\psi_b^c|H_{opt}|\psi_b^c\rangle\right) = C_1(T_1 - T_1) = 0 \quad (S4)$$

For transitions between symmetric bands we have (to first order)

$$\langle\psi_a^{v\prime}|H_{opt}|\psi_a^{c\prime}\rangle = \langle\psi_a^v|H_{opt}|\psi_a^c\rangle = T_2$$

$$\langle\psi_b^{v\prime}|H_{opt}|\psi_b^{c\prime}\rangle = \langle\psi_b^v|H_{opt}|\psi_b^c\rangle = -T_2$$

Therefore we obtain the selection rule that in region I optical transitions only happen between symmetric bands, which give little optical conductivity change compared to bare graphene. We have not assumed any specific form of Moiré potential, therefore this conclusion is valid for all types of effective potential.

## S4. Selection rules for optical transitions near sBZ boundary

From above discussion, we see that the optical conductivity change is mainly coming from states near sBZ boundary, where both band dispersion and electron wavefunction are strongly modified.

The gaps at m points result from the electronic coupling between originally degenerate states $\psi_1, \psi_2$ due to Moiré potential (See Fig. S2b). The electronic coupling matrix element is $i(u_3 \mp u_1)$ for the conduction/valence band, and the resulting energy gap will be $|2(u_3 \mp u_1)|$ based on the first order degenerate perturbation theory.

Note that the coupling matrix elements are the same for conduction and valence band with pseudospin-dependent potential $u_3$; but they are opposite with pseudospin-mixing potential $u_1$. This difference between $u_3$ and $u_1$ potential leads to distinctly different optical transition selection rule.

In the case of pseudospin-dependent potential $u_3$, the wavefunctions of states 1e, 2h, 1h, 2h at m point are:

$$\psi_{2e} = (\psi_1^c + i\psi_2^c)/\sqrt{2}, \qquad \psi_{1e} = (\psi_1^c - i\psi_2^c)/\sqrt{2},$$

$$\psi_{1h} = (\psi_1^v + i\psi_2^v)/\sqrt{2}, \qquad \psi_{2h} = (\psi_1^v - i\psi_2^v)/\sqrt{2}$$

For both conduction and valence sides, the higher energy bands (2e and 1h) have the same "+" sign due to the same coupling matrix elements. Similar to equation S3, for unperturbed graphene electronic states the optical matrix element is:

$$\langle\psi_1^v|H_{opt}|\psi_2^v\rangle = \langle\psi_1^v|H_{opt}|\psi_2^c\rangle = \langle\psi_1^c|H_{opt}|\psi_2^v\rangle = -\langle\psi_1^c|H_{opt}|\psi_2^c\rangle = 0$$

$$\langle\psi_1^v|H_{opt}|\psi_1^c\rangle = \langle\psi_2^c|H_{opt}|\psi_2^v\rangle = -\langle\psi_1^c|H_{opt}|\psi_1^v\rangle = -\langle\psi_2^v|H_{opt}|\psi_2^c\rangle = iv_F\hat{\mathbf{z}}\cdot(\mathbf{E}\times\hat{\mathbf{k}}_1) = T_2$$

The optical transition matrix element for perturbed states

$$\langle\psi_{1h}|H_{opt}|\psi_{2e}\rangle = \langle\psi_{2h}|H_{opt}|\psi_{1e}\rangle = (\langle\psi_1^v|H_{opt}|\psi_1^c\rangle + \langle\psi_2^v|H_{opt}|\psi_2^c\rangle)/2 = 0$$

$$\langle\psi_{1h}|H_{opt}|\psi_{1e}\rangle = \langle\psi_{2h}|H_{opt}|\psi_{2e}\rangle = (\langle\psi_1^v|H_{opt}|\psi_1^c\rangle - \langle\psi_2^v|H_{opt}|\psi_2^c\rangle)/2 = T_2$$

Therefore only transitions between 1h-1e and 2h-2e are allowed (Fig. 3c).

In the case of pseudospin-mixing potential $u_1$, we have:

$$\psi_{2e} = (\psi_1^c - i\psi_2^c)/\sqrt{2}, \quad \psi_{1e} = (\psi_1^c + i\psi_2^c)/\sqrt{2},$$

$$\psi_{1h} = (\psi_1^v + i\psi_2^v)/\sqrt{2}, \quad \psi_{2h} = (\psi_1^v - i\psi_2^v)/\sqrt{2}$$

The optical transition matrix element

$$\langle\psi_{1h}|H_{opt}|\psi_{1e}\rangle = \langle\psi_{2h}|H_{opt}|\psi_{2e}\rangle = (\langle\psi_1^v|H_{opt}|\psi_1^c\rangle + \langle\psi_2^v|H_{opt}|\psi_2^c\rangle)/2 = 0$$

$$\langle\psi_{1h}|H_{opt}|\psi_{2e}\rangle = \langle\psi_{2h}|H_{opt}|\psi_{1e}\rangle = (\langle\psi_1^v|H_{opt}|\psi_1^c\rangle - \langle\psi_2^v|H_{opt}|\psi_2^c\rangle)/2 = T_2$$

Therefore, only transitions between 1h-2e and 2h-1e are allowed (Fig. 3d).

We notice that, the two types of gaps here, the "inverse" gap and "normal" gap, gives an interesting analogy to topological insulator and normal insulator. Our gap equation $u_3 \mp u_1$ is very similar to the gap equation $\Delta_3 \mp \Delta_1$ obtained in the four band topological insulator model by Kane and Mele[5]. In that model, topological phase transition happens at $\Delta_3 = \Delta_1$, which is analogous to the transition between "inverse" and "normal" gaps in our case. And this transition corresponds to a change from absorption peak to dip at $E_M$ in optical spectra. Another consequence from the gap equation $u_3 \mp u_1$ is the asymmetry between electron and hole bands. The valence (conduction) band will be modified more strongly if $u_3$ and $u_1$ have the same (opposite) sign. We observe a much stronger modification on hole side (see Fig.

1c), consistent with previous studies[6-10]. This can be explained naturally by the existence of both $u_3$ and $u_1$ with the same sign.

## S5. Microscopic model to determine $u_0$, $u_3$ and $u_1$

Ref [11] and [12] used a graphene/BN hopping model and obtained the parameters for zero twist-angle graphene/BN system:

$$V_0(u_0, u_3, u_1) = \left(\frac{|M_{B-C}|^2}{E_B - E_g} + \frac{|M_{N-C}|^2}{E_N - E_g}\right)\left(\frac{1}{2}, -\frac{\sqrt{3}}{2}, -1\right) = V_S\left(\frac{1}{2}, -\frac{\sqrt{3}}{2}, -1\right)$$

where $M_{B-C}/M_{N-C}$ are the coupling matrix element between graphene and boron/nitrogen states, respectively, which can be determined by the pair interaction potential between graphene and the substrate atoms; $E_g$ is the energy of graphene state; $E_B = E_g + 3.1eV$ and $E_N = E_g - 1.5\ eV$ are the energy of the BN conduction and valence states at the K point, and they are localized at the boron and nitrogen lattice sites, respectively[13]. We see that the sign of $V_S$ is determined by whether graphene couples stronger to boron states or nitrogen states.

Ref. [11] assumes that both carbon-boron and carbon-nitrogen interaction have the same strength as carbon-carbon interaction in bilayer graphene, and obtains a $V_S \sim -3\text{meV}$. However, the pair interaction strength is sensitive on the atomic orbital radius, which is different for boron, carbon and nitrogen. The p orbital radii of boron, carbon and nitrogen atoms are roughly 80, 60, and 50 pm, respectively. The significantly larger radius of boron p orbital indicates a larger overlap and therefore stronger pair interaction of carbon-boron than carbon-carbon; and the opposite case for carbon-nitrogen. We roughly estimate the effective potential strength $V_B = 10\text{meV}$ and $V_N = -1\text{meV}$. The latter is much weaker and can be neglected.

Based on these evaluations, we obtain the spinor effective potential $V_0 = 10\text{meV}$ and $(u_0, u_3, u_1) = \left(\frac{1}{2}, -\frac{\sqrt{3}}{2}, -1\right)$.

**Reference:**

1. Geick, R., Perry, C. H. & Rupprech.G NORMAL MODES IN HEXAGONAL BORON NITRIDE. *Physical Review* **146**, 543-& (1966).
2. Wang, F. *et al.* Gate-variable optical transitions in graphene. *Science* **320**, 206-209 (2008).
3. Li, Z. Q. *et al.* Dirac charge dynamics in graphene by infrared spectroscopy. *Nature Physics* **4**, 532-535 (2008).
4. Nair, R. R. *et al.* Fine structure constant defines visual transparency of graphene. *Science* **320**, 1308-1308 (2008).
5. Kane, C. L. & Mele, E. J. Quantum Spin Hall Effect in Graphene. *Physical Review Letters* **95**, 226801 (2005).
6. Yankowitz, M. *et al.* Emergence of superlattice Dirac points in graphene on hexagonal boron nitride. *Nature Physics* **8**, 382-386 (2012).
7. Ponomarenko, L. A. *et al.* Cloning of Dirac fermions in graphene superlattices. *Nature* **497**, 594-597 (2013).
8. Dean, C. R. *et al.* Hofstadter's butterfly and the fractal quantum Hall effect in moire superlattices. *Nature* **497**, 598-602 (2013).
9. Hunt, B. *et al.* Massive Dirac fermions and Hofstadter butterfly in a van der Waals heterostructure. *Science* **340**, 1427-1430 (2013).
10. Yang, W. *et al.* Epitaxial growth of single-domain graphene on hexagonal boron nitride. *Nature Materials* **12**, 792-797 (2013).
11. Kindermann, M., Uchoa, B. & Miller, D. L. Zero-energy modes and gate-tunable gap in graphene on hexagonal boron nitride. *Physical Review B* **86** (2012).
12. Wallbank, J. R. *et al.* Generic miniband structure of graphene on a hexagonal substrate. *Physical Review B* **87** (2013).
13. Slawinska, J., Zasada, I. & Klusek, Z. Energy gap tuning in graphene on hexagonal boron nitride bilayer system. *Physical Review B* **81** (2010).